\documentclass[aps,pra,twocolumn,amsmath,amssymb]{revtex4-2}

\usepackage{graphicx}
\usepackage{color}
\usepackage[hidelinks]{hyperref}

\hypersetup{colorlinks=true, linkcolor=black, citecolor=black, urlcolor=blue}

\begin{document}
\author{Karol Gietka}
\email[Corresponding author: ]{karol.gietka@oist.jp}

\affiliation{Quantum Systems Unit, Okinawa Institute of Science and Technology Graduate University, Onna, Okinawa 904-0495, Japan} 

\title{Harnessing the center-of-mass excitations in quantum metrology}

%%%%%%%%%%%%%%%%%%%%%%%%%%%%%%%%%%%%%%%%%%%%%%%%%%%%%%%%%%%%%%%%%%%%%%%%%%%%
%%%%%%%%%%%%%%%%%%%%%%%%%%%%%%%  ABSTRACT  %%%%%%%%%%%%%%%%%%%%%%%%%%%%%%%%%
%%%%%%%%%%%%%%%%%%%%%%%%%%%%%%%%%%%%%%%%%%%%%%%%%%%%%%%%%%%%%%%%%%%%%%%%%%%%

\begin{abstract}
In quantum metrology, one typically creates correlations among atoms or photons to enhance measurement precision. Here, we show how one can use other excitations to perform quantum-enhanced measurements on the example of center-of-mass excitations of a spin-orbit coupled Bose-Einstein condensate and a Coulomb crystal. We also present a method to simulate a homodyne detection of center-of-mass excitations in these systems, which is required for optimal estimation.
\end{abstract}
\date{\today}
\maketitle

%%%%%%%%%%%%%%%%%%%%%%%%%%%%%%%%%%%%%%%%%%%%%%%%%%%%%%%%%%%%%%%%%%%%%%%%%%%%
%%%%%%%%%%%%%%%%%%%%%%%%%%%%%%%  INTRODUCTION  %%%%%%%%%%%%%%%%%%%%%%%%%%%%%
%%%%%%%%%%%%%%%%%%%%%%%%%%%%%%%%%%%%%%%%%%%%%%%%%%%%%%%%%%%%%%%%%%%%%%%%%%%

\section{Introduction}
Quantum metrology relies on exploiting non-classical correlations between system constituents to overcome the standard quantum limit of precision when estimating an unknown parameter~\cite{2006Quantum_Metrology,2011advancesinQM}. In typical situations, these correlations are being created among atoms~\cite{2019_rmp_qmetrology} or photons~\cite{2015DEMKOWICZDOBRZANSKI2015345,2020photonic_QM} which are subsequently used to imprint the information about the unknown parameter. Then, the final state of the system is measured, and the resulting measurement outcomes are used to estimate the parameter. Unfortunately, in some cases, obtaining a suitable correlated state of atoms or photons might be challenging, hindering the possibility of performing quantum-enhanced measurements. However, atoms and photons do not exhaust the set of possible excitations among which the correlations could be generated and potentially used~\cite{2015squeezingmechanicalresonantor,2018phononNOON,2021rapidsqueezingmotion,2022mechanicalsqueezing,2022entangledmechanicalreseonateors}. 

Let us, for example, consider a single atom in a harmonic trap for which it seems impossible to create quantum correlations for metrological purposes. Nevertheless, if the harmonic trap has a very high frequency, the wavefunction of the atom will be very localized, which means that the center-of-mass excitations must be correlated (squeezed). Such squeezing is known to enhance the precision of measurement in the context of photons~\cite{2019quantumsensingwithsqueezedlight}; unfortunately, most of the forces that one would want to measure do not couple to the center-of-mass motion (except gravity), but to the atomic internal degrees of freedom, thus such correlations might seem indeed metrologically useless.

To overcome this apparent limitation, one could couple the center-of-mass motion of an atom with its internal degree of freedom (for example, via spin-orbit coupling) and potentially transfer the information from an atom whose internal structure is affected by some force to the center-of-mass motion and subsequently measure the position or momentum distribution to estimate the unknown parameter. Then the precision of such estimation should scale with the number of center-of-mass excitations which for a very squeezed state might be very large even for a single atom. A similar approach is typically exploited in light-matter systems. There, some force affects an ensemble of atoms, and due to the light-matter interactions, the state of the light is also affected by that force~\cite{2019supersolid_Gravimeter_Gietka,2021_Gietka_cavity_magnetometer,2021_Amor_s_Binefa_NJP}. Performing then quadrature measurements on the light usually allows one to estimate the strength of the force. In this case, the precision scales typically with the number of field excitations (photons).

In this manuscript, on the example of center-of-mass excitations, we show how one can harness less obvious excitations in quantum-enhanced measurements in analogy to light-matter systems. To this end, we consider an adiabatic protocol involving a spin-orbit coupled Bose-Einstein condensate and a (sudden) quench protocol involving a Coulomb crystal. We also show how one can simulate a homodyne detection of center-of-mass excitations in these systems which is required for optimal estimation.

%%%%%%%%%%%%%%%%%%%%%%%%%%%%%%%%%%%%%%%%%%%%%%%%%%%%%%%%%%%%%%%%%%%%%%%%%%%%
%%%%%%%%%%%%%%%%%%%%%%%%%%%%%%%  SOC BEC  %%%%%%%%%%%%%%%%%%%%%%%%%%%%%%%%%%
%%%%%%%%%%%%%%%%%%%%%%%%%%%%%%%%%%%%%%%%%%%%%%%%%%%%%%%%%%%%%%%%%%%%%%%%%%%%

\section{Adiabatic protocol with a spin-orbit coupled Bose-Einstein condensate}
A spin-orbit coupled Bose-Einstein condensate is a system in which atoms are coupled to different momenta depending on their spin state. Although it is well understood~\cite{2016propertiessocbec} and has been realized experimentally~\cite{2011socbec,2016socrealizationketterle}, its metrological potential remains unexplored. In the simplest possible case of a two-mode non-interacting condensate, the behavior of each atom can be described by the following single-particle Hamiltonian (we set $\hbar=1$ throughout the entire manuscript)
\begin{align}\label{eq:H1}
    \hat H = \frac{ \left(\hat p + k \hat \sigma_z\right)^2}{2m} + \frac{m \omega^2 \hat x^2}{2} + \frac{ \Omega}{2}  \hat \sigma_x
\end{align}
where $\hat x$ and $\hat p \equiv - i \partial_x$ are position and momentum operators of an atom, $\omega$ is the frequency of the harmonic trap, $m$ is the atomic mass, $\Omega$ is the atomic frequency, $\hat \sigma_i$ are the standard Pauli matrices, and $k$ is the spin-orbit interaction strength. Depending on the strength of the coupling between the two modes of the condensate $\Omega$, one can distinguish two phases. The single minimum phase is characterized by $\Omega > 2k^2/m$, and a double minimum phase (stripe phase) is characterized by $\Omega < 2k^2/m$.

For the sake of comparison with light-matter systems, we can now rotate Hamiltonian \eqref{eq:H1} and rewrite it in the form of a quantum Rabi model using the annihilation and creation operators of the harmonic oscillator 
\begin{align}\label{eq:qrform}
     \hat H =  \omega \hat a^\dagger \hat a + \frac{\Omega}{2} \hat \sigma_z + k\sqrt{\frac{ \omega}{2m}} (\hat a +  \hat a^\dagger) \hat \sigma_x.
\end{align}
In other words, a non-interacting spin-orbit coupled Bose-Einstein condensate of $N$ atoms trapped in a harmonic potential is equivalent to $N$ indistinguishable copies of a quantum Rabi model with critical coupling strength $k_c = \sqrt{\Omega m/2}$. The latter has been extensively studied in the context of light-matter interactions~\cite{2012qrmsimulationLM,2016rabimodel80} and (critical) quantum metrology~\cite{2020CriticalParis,2021Dyanmicalframework,2022critialcontinous,2022criticalspeedup}.

%%%%%%%%%%%%%%%%%%%%%%%%%%%%%%%%%%%%%%%%%%%%%%%%%%%%%%%%%%%%%%%%%%%%%%%%%%%%
%%%%%%%%%%%%%%%%%%%%%%%%%%%%%%%  EQUILIBRIUM  %%%%%%%%%%%%%%%%%%%%%%%%%%%%%%
%%%%%%%%%%%%%%%%%%%%%%%%%%%%%%%%%%%%%%%%%%%%%%%%%%%%%%%%%%%%%%%%%%%%%%%%%%%%

Let us now discuss an equilibrium protocol which relies on adiabatic preparation of the ground state of the system in the single minimum phase~\cite{2020CriticalParis}. For the sake of brevity, we assume now that $\Omega$ is an unknown parameter, and we can control all other parameters to an arbitrary extent. In order to find the ground state (in a suitable form), we can apply the Schrieffer-Wolff transformation~\cite{1966SWtrans}
\begin{align}
    \hat U =\exp\left(i \frac{k}{k_c}\frac{\sqrt{\omega}}{2\sqrt{\Omega}}(\hat a +\hat a^\dagger){\hat \sigma_y} \right), 
\end{align}
and for $  1-k^2/k_c^2\gg({\omega}/{\Omega})^{2/3} $ rewrite the Hamiltonian as (see Appendix~\ref{app:effsocbec} for details)
\begin{align} \label{eq:effsocbec}
     \hat H \approx \omega \hat a^\dagger \hat a + \frac{ \Omega}{2} \hat \sigma_z + k^2{\frac{  \omega}{2m \Omega}} (\hat a +  \hat a^\dagger)^2 \hat \sigma_z.
\end{align}
whose ground state is a squeezed vacuum coupled to the spin down state
$
    |{\psi} \rangle = \hat S(\xi) |0\rangle \otimes|\!\!\downarrow\,\rangle
$ with $\hat S(\xi) \equiv \exp\{(\xi/2)(\hat a^\dagger)^2-(\xi^*/2)\hat a^2\}$ being the squeeze operator with $\xi = -\frac{1}{4} \ln\{1-(k/k_c)^2\}$. We can now use this ground state to calculate the quantum Fisher information $\mathcal{I}_\Omega = \langle \partial_\Omega \psi|\partial_\Omega \psi\rangle - \langle \psi|\partial_\Omega \psi \rangle^2$ which is related to the sensitivity of estimating an unknown parameter through the Cram\'er-Rao bound $\Delta^2\Omega \geq 1/ N \mathcal{I}_\Omega$~\cite{2009pezzesmerzientHL} with $N$ being the number of system's copies. The quantum Fisher information becomes
\begin{align}
    \mathcal{I}_\Omega= \frac{1}{8\Omega^2\left(1-\frac{k^2}{k_c^2}\right)^2}\frac{k^4}{k_c^4}.
\end{align}
Since the ground state is a squeezed vacuum state, which is entirely characterized by the width of the distribution (a consequence of Gaussianity), the measurement that leads to such Fisher information is the measurement of the second moment of position (or momentum)~\cite{2009qstomographyRMP,2012gaussianrmp}. The above expression tells us, in fact, how distinguishable neighboring ground states (as a function of $k$) are~\cite{1994distancestates}. In order to make a metrological sense out of this expression, we now assume that the coupling was adiabatically changed from $k=0$ to $k=k_f< k_c$ which took time~\cite{2020CriticalParis}
$
    T_{k_f}\approx\left(2\gamma\omega \sqrt{1-{k_f^2}/{k_c^2}}\right)^{-1}
$
where $\gamma\ll1$. Here we see the effect of critical slowing down; increasing $k$ towards $k_c$ closes the energy gap which means that preparation time also has to increase. Plugging this time into the expression for the quantum Fisher information yields
\begin{align}
    \mathcal{I}_\Omega = 8 \frac{\omega^2}{\Omega^2}\gamma^2\langle \hat n\rangle^2 T_{k_f}^2,
\end{align}
where
$
\langle \hat n\rangle= \sinh^2\xi \approx \left(4\sqrt{1-{k_f^2}/{k_c^2}}\right)^{-1}
$ is the average number of center-of-mass excitations~\cite{1983squeezedstates}. Thus, the Cram\'er-Rao bound becomes
\begin{align}
    \Delta^2 \Omega \geq \left(8 N \frac{\omega^2}{\Omega^2} \gamma^2\langle \hat n\rangle^2 T_{k_f}^2 \right)^{-1}
\end{align}
and exhibits a Heisenberg scaling with respect to the center-of-mass excitations (similar scalings can be shown for other parameters of the system).

\begin{figure}[htb!]
    \centering
    \includegraphics[width=0.48\textwidth]{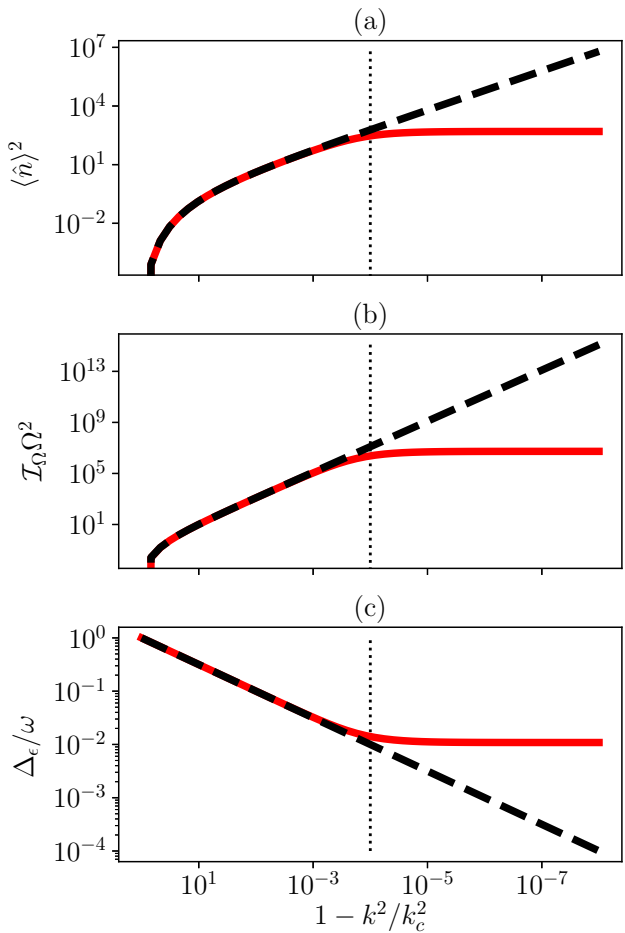}
    \caption{Comparison between the effective model (dashed black line) and Hamiltonian~\eqref{eq:H1} (solid red line) for $\Omega/\omega = 10^6$. The vertical dotted line indicates $ 1-k^2/k_c^2 = ({\omega}/{\Omega})^{2/3}$. (a), (b), and (c) shows the number of excitations, quantum Fisher information, and energy gap, respectively.}
    \label{fig:fig1}
\end{figure}

The above expression for the quantum Fisher information (and Cram\'er-Rao bound) is valid as long as $ 1-k^2/k_c^2\gg({\omega}/{\Omega})^{2/3}$; in order to check what happens once this condition is no longer satisfied, we have to perform numerical calculations. These are presented in Fig.~\ref{fig:fig1} where we calculate the number of excitations, quantum Fisher information, and the normalized energy gap $\Delta_\epsilon/\omega$ as a function of $1-k^2/k_c^2$ for a fixed $\omega/\Omega$ using the Hamiltonian~\eqref{eq:H1} (see also Appendix~\ref{app:effsocbec}).
As can be seen in Fig.~\ref{fig:fig1}, once the condition $1-k^2/k_c^2\gg({\omega}/{\Omega})^{2/3}$ is no longer satisfied, the state does not change anymore and the number of excitations (Fig.~\ref{fig:fig1}a), quantum Fisher information (Fig.~\ref{fig:fig1}b), and the energy gap (Fig.~\ref{fig:fig1}c) does not alter.

The above results prove that adiabatic preparation of the ground-state of a spin-orbit coupled Bose-Einstein condensate can, in principle, lead to sensitivities exhibiting Heisenberg scaling with the number of center-of-mass excitations without exploiting any correlations between the atoms. In practice, however, overcoming the standard quantum limit with an adiabatic approach might be substantially limited due to the critical slowing down effect. This is caused by the fact that during adiabatic dynamics, the (squeezed) excitations are created very slowly---and thus ineffectively---as we are closing the energy gap while increasing $k$ towards $k_c$. In order to remove the problem of critical slowing down, one can perform a fast quench beyond the transition point $k_c$ to increase the rate at which the excitations are being created and thus make an advantage out of closing energy gap. Although one can apply the sudden quench protocol to the spin-orbit coupled Bose-Einstein condensate, in order to show the generality of using center-of-mass excitations, we will now demonstrate this mechanism on the example of a Coulomb crystal.

%%%%%%%%%%%%%%%%%%%%%%%%%%%%%%%%%%%%%%%%%%%%%%%%%%%%%%%%%%%%%%%%%%%%%%%%%%%%
%%%%%%%%%%%%%%%%%%%%%%%%%%%%%%%%  DYNAMIC  %%%%%%%%%%%%%%%%%%%%%%%%%%%%%%%%%
%%%%%%%%%%%%%%%%%%%%%%%%%%%%%%%%%%%%%%%%%%%%%%%%%%%%%%%%%%%%%%%%%%%%%%%%%%%%

\section{Sudden quench protocol with a Coulomb crystal}
A Coulomb crystal~\cite{2015ionCoulombCrystal} is a crystal-like system composed of cold ions which can serve as a quantum simulator. For example, in a recent experiment~\cite{2019CrystalDickesym} a Coulomb crystal in the presence of an external transverse field was used to engineer a simulator of the paradigmatic Dicke Hamiltonian~\cite{2011DickeModelreview}. This was achieved by coupling vibrational modes of the crystal with the spin degree of freedom via a spin-dependent optical dipole force generated by the interference of a pair of lasers. In the presence of an additional transverse field, the system is described by the Dicke Hamiltonian
\begin{equation}\label{eq:culcrys}
    \hat H = - \delta \hat a^\dagger \hat a + B\hat S_x -  \frac{g}{\sqrt{N}}(\hat a + \hat a^\dagger)\hat S_z,
\end{equation}
where $\hat a$ ($\hat a^\dagger$) is the annihilation (creation) operator of the center-of-mass excitation and $\hat S_i$ are the collective spin operators. In the above Hamiltonian $\delta \equiv \omega_R-\omega$ is detuning between the laser frequency $\omega_R$ and the center-of-mass mode frequency $\omega$, $g$ is the homogeneous coupling between each ion and the center-of-mass mode, and $B$ is the strength of the applied transverse field. Depending on the transverse field strength, one can distinguish two phases. The paramagnetic (normal) phase for $B>g^2/|\delta|\equiv B_c$ and the ferromagnetic (superradiant) phase for $B<B_c$.

Dynamic metrological protocol relies on performing a sudden quench from the single minimum (normal) phase to the double minimum (superradiant) phase and~\cite{Hayes_2018} exploits the dynamics of an inverted harmonic oscillator~\cite{2021Inverted_Oscillator_Gietka}. This provides an exponential speed up with respect to the adiabatic protocol~\cite{2021understandingCQM}. In the following, we treat $\delta$ as the unknown parameter and change $B$ to perform a quench between the two phases. In order to see the effect of an inverted harmonic oscillator, we first apply the Schrieffer-Wolff transformation 
\begin{equation}
    \hat U =  \exp\left(i \frac{B_c}{B }\frac{\delta}{g\sqrt{N}}(\hat a +\hat a^\dagger){\hat S_y} \right), 
\end{equation}
to the Hamiltonian~\eqref{eq:culcrys} and rewrite it with the help of quadrature operators~(see Appendix~\ref{app:effcoucrys})
\begin{equation}\label{eq:cryseff}
    \hat H = \frac{\delta}{2}\hat P^2 + \frac{\delta}{2}\left(1-\frac{B_c}{B}\right)\hat X^2.
\end{equation}
The above Hamiltonian describes a harmonic oscillator with a tunable frequency which can become negative. It gives an accurate ground state description only in the paramagnetic phase ($B>B_c$), as for $B<B_c$ the ground state does not exist. Nevertheless, this Hamiltonian can be still used to describe quenches from paramagnetic to the ferromagnetic phase until a certain number of center-of-mass excitations is reached ($n_\mathrm{max}\propto N B/\delta$), and we assume now that this number is not exceeded (see Appendix~\ref{app:effcoucrys}).
 \begin{figure}[htb!]
    \centering
    \includegraphics[width=0.48\textwidth]{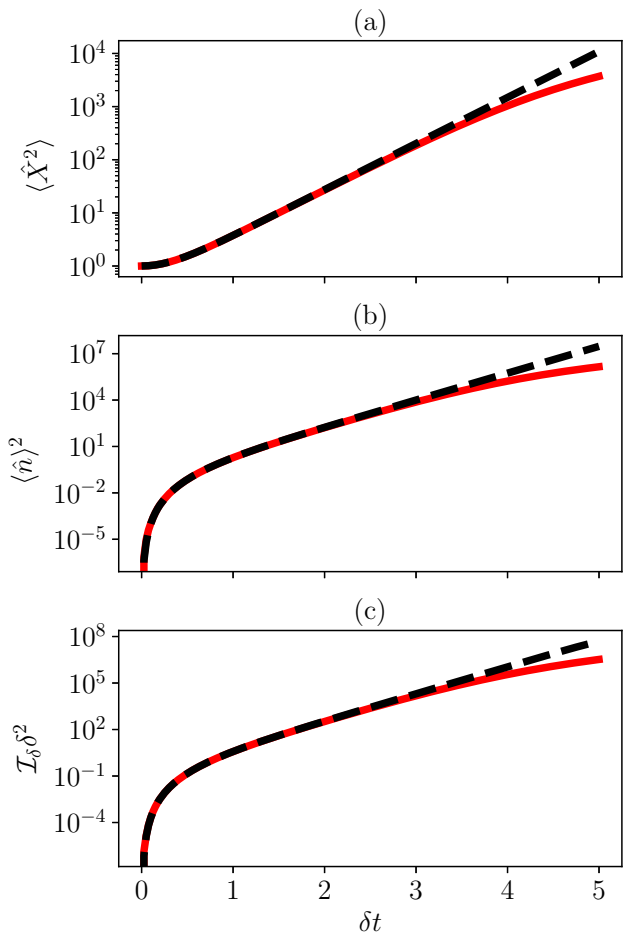}
    \caption{Comparison between the effective model~\eqref{eq:cryseff} (dashed-black line) and the Hamiltonian~\eqref{eq:culcrys} (solid red line) for $B_f =B_c/2$. (a), (b), and (c) shows $\langle \hat X^2\rangle$, number of excitations, and normalized quantum Fisher information, respectively. For the numerical calculations, we have set $N=1$, $B=B_c/2$, $g/\delta = 80$, and $B/\delta = 3200$.}
    \label{fig:fig2}
\end{figure}

In order to calculate the quantum Fisher information, we assume that the system is quenched from the ground state at $B\gg B_c$, where the center-of-mass excitations are decoupled from the spin, to some $B_f<B_c$. Unfortunately, apart from dependence on hyperbolic functions (see Appendix~\ref{app:qficrys}) which reflects the fact that inverted harmonic oscillator dynamics stimulates squeezing, the full expression for quantum Fisher information is complicated and one does not clearly see its advantage. For a special case of $B_f=B_c/2$, one can show that the quantum Fisher information is equal to
\begin{equation}
    \mathcal{I}_\delta = \frac{B_c^2}{2 \delta^2 B_f^2}\sinh^4 \delta t =  \frac{2}{ \delta^2 }\langle \hat n \rangle^2,
 \end{equation}
where $\langle \hat n \rangle = \sinh^2 \delta t$ is the instantaneous number of center-of-mass excitations.

In order to inspect what happens once the inverted harmonic oscillator dynamics is no longer valid, we have to resort to numerical calculations. These are presented in Fig.~\ref{fig:fig2}, where we can see $\langle \hat X^2 \rangle$, $\langle \hat n\rangle^2$, and $\mathcal{I}_\delta$ as a function of time after the quench. As can be seen in Fig.~\ref{fig:fig2}(b), once the number of excitations does not grow exponentially, the same happens with the quantum Fisher information [Fig.~\ref{fig:fig2}(c)] and squeezing [Fig.~\ref{fig:fig2}(a)].

In the case of a quench protocol, the final state is Gaussian, and the optimal measurement that can lead to saturation of the Cram\'er-Rao bound is also a homodyne detection but at an angle that is dependent on time and parameters of the system~\cite{2021understandingCQM}. Therefore, one needs a method to simulate a homodyne detection at an arbitrary angle for the center-of-mass excitations.

% %%%%%%%%%%%%%%%%%%%%%%%%%%%%%%%%%%%%%%%%%%%%%%%%%%%%%%%%%%%%%%%%%%%%%%%%%%%%
% %%%%%%%%%%%%%%%%%%%%%%%%%%%%%%%  HOMODYNE %%%%%%%%%%%%%%%%%%%%%%%%%%%%%%%%%%
% %%%%%%%%%%%%%%%%%%%%%%%%%%%%%%%%%%%%%%%%%%%%%%%%%%%%%%%%%%%%%%%%%%%%%%%%%%%%

\section{Simulating homodyne detection}
In the case of light-matter systems and Gaussian states of light, it is well known that in order to extract the maximum amount of information from the system one has to perform a homodyne measurement in the optimal direction~\cite{2006optimalgaussianmeasure}, which can be controlled with the phase of the local oscillator. For a Bose-Einstein condensate and a Coulomb crystal, it seems that one is limited to the quadrature measurement in only two directions, namely $\hat x$ ($\hat X$) and $\hat p$ ($\hat P$). In the following, we will explain how an equivalent of a generalized quadrature operator $\hat Q(\theta) = \hat x \cos \theta  + \hat p \sin \theta $ ($\theta$ is the local oscillator phase) for the center-of-mass excitations could be measured in a spin-orbit coupled Bose-Einstein condensate and a Coulomb crystal. To this end, let us consider Hamiltonian~\eqref{eq:H1} for $k=0$. Such a Hamiltonian describes two subsystems that are decoupled. This means that for $k=0$, the dynamics of center-of-mass excitations is governed solely by $\hat H = \omega \hat a^\dagger \hat a$. Such a Hamiltonian generates rotations in the quadrature phase space around its origin. Therefore, by performing a sudden quench back to $k=0$ after the final state was reached, it is possible to rotate the wavefunction by an arbitrary angle $\theta$ in the phase space and perform thus a measurement of $\hat Q(\theta)$ by measuring either momentum or position distribution. 

In the case of a Coulomb crystal [Hamiltonian~\eqref{eq:culcrys}], decoupling of the collective spin from the center-of-mass motion can be done by turning off the lasers or, alternatively, quenching the magnetic field back to $B\gg B_c$ after the final state is obtained. Therefore an arbitrary rotation can be performed in the phase space of center-of-mass excitations by exploiting an appropriate parameter quench. Then $\hat Q(\theta)$ could be measured by inspecting either momentum or position distribution (see Appendix~\ref{app:homodyne} for a visualization) similarly as in the case of a spin-orbit-coupled Bose-Einstein condensate.

%%%%%%%%%%%%%%%%%%%%%%%%%%%%%%%%%%%%%%%%%%%%%%%%%%%%%%%%%%%%%%%%%%%%%%%%%%%%
%%%%%%%%%%%%%%%%%%%%%%%%%%%%%%  CONCLUSIONS  %%%%%%%%%%%%%%%%%%%%%%%%%%%%%%%
%%%%%%%%%%%%%%%%%%%%%%%%%%%%%%%%%%%%%%%%%%%%%%%%%%%%%%%%%%%%%%%%%%%%%%%%%%%%
\section{Conclusions}
In this work, we have shown how to exploit excitations other than photonic or atomic to enhance measurement precision. To this end, we have proposed how to exploit center-of-mass excitations of a Bose-Einstein condensate and a Coulomb crystal. We presented two protocols based on an adiabatic and a sudden quench, originally developed for light-matter systems, that could be used to estimate unknown parameters of the system with sensitivities exhibiting Heisenberg scaling with respect to the center-of-mass excitations. We have also presented how one can simulate the measurement of a generalized quadrature operator, which is required for extracting the maximal amount of information about the unknown parameter from these systems. This could be achieved by decoupling the center-of-mass mode from the spin degree of freedom, waiting until the state \emph{aligns} with the position or momentum axis, and finally measuring its distribution. Then, as the resulting states are Gaussian, the second moment of position and momentum can serve as the optimal estimator.

In the case of a spin-orbit coupled Bose-Einstien condensate, we did not consider any interactions. In general, small spin-spin repulsive interactions will shift the critical spin-orbit coupling strength to a larger value and \emph{change the shape} of the wavefunction (it will no longer be Gaussian). This, however,  should not prevent using center-of-mass excitations in quantum metrology. The major difference will be the necessity to use the entire distribution of position (or momentum) to construct an optimal estimator. 

In both cases, for finite temperature systems, where the system occupies incoherently excited states, it can be shown that while the effective Hamiltonian description is valid, the quantum Fisher information increases with growing temperature (see Appendix~\ref{app:finiteTBEC}) which is caused by the increased number of excitations in a thermal state. Once the effective harmonic oscillator model is not valid anymore, the quantum Fisher information decreases with increasing temperature.  In this case, the entire distribution of position (or momentum) is also required for optimal estimation as the state is no longer Gaussian.

The presented protocols could be used in any system that can be mapped to a harmonic oscillator coupled to a (collective) spin; one intriguing example is electrons trapped on the surface of liquid Helium, where the role of the harmonic oscillator is played by Landau levels~\cite{2019heliumrabi}. We hope that this work will stimulate the search for new metrological protocols (in known and well-understood systems) involving various kinds of excitations that might be more robust to decoherence and losses than atoms or photons. In particular, protocols involving spin-orbit coupled fermionic atoms~\cite{tobe2022} (or a Tonks-Girardeau gas~\cite{1936tonks,1963girardeau,2004tgEXP}) might lead to protocols robust to atom losses and benefiting from both correlated center-of-mass excitations and naturally existing fermionic correlations~\cite{2014marzozlinfermionmet}.

%%%%%%%%%%%%%%%%%%%%%%%%%%%%%%%%%%%%%%%%%%%%%%%%%%%%%%%%%%%%%%%%%%%%%%%%%%%%
%%%%%%%%%%%%%%%%%%%%%%%%%%%% ACKNOWLEDGEMENTS  %%%%%%%%%%%%%%%%%%%%%%%%%%%%%
%%%%%%%%%%%%%%%%%%%%%%%%%%%%%%%%%%%%%%%%%%%%%%%%%%%%%%%%%%%%%%%%%%%%%%%%%%%%

\section{Acknowledgements}
K.G. is pleased to acknowledge Mohamed Hatifi, Thom\'as Fogarty, Ayaka Usui, Yongping Zhang, and Thomas Busch for fruitful discussions. Simulations were performed using the open-source \textsc{QuantumOptics.jl} framework in \textsc{Julia}~\cite{kramer2018quantumoptics}. This work was supported by the Okinawa Institute of Science and Technology Graduate University.

%%%%%%%%%%%%%%%%%%%%%%%%%%%%%%%%%%%%%%%%%%%%%%%%%%%%%%%%%%%%%%%%%%%%%%%%%%%%
%%%%%%%%%%%%%%%%%%%%%%%%%%%%%%%% APPENDIX  %%%%%%%%%%%%%%%%%%%%%%%%%%%%%%%%%
%%%%%%%%%%%%%%%%%%%%%%%%%%%%%%%%%%%%%%%%%%%%%%%%%%%%%%%%%%%%%%%%%%%%%%%%%%%%

\onecolumngrid
 \appendix
 
 %%%%%%%%%%%%%%%%%%%%%%%%%%%%%%%%%%%%%%%%%%%%%%%%%%%%%%%%%%%%%%%%%%%%%%%%%%%%
%%%%%%%%%%%%%%%%%%%%%%%%%%%%%%%% APPENDIX A %%%%%%%%%%%%%%%%%%%%%%%%%%%%%%%%%
%%%%%%%%%%%%%%%%%%%%%%%%%%%%%%%%%%%%%%%%%%%%%%%%%%%%%%%%%%%%%%%%%%%%%%%%%%%%

 \section{Effective Hamiltonian for the spin-orbit coupled Bose-Einstein condensate} \label{app:effsocbec}
 Here we derive the effective Hamiltonian~\eqref{eq:effsocbec}. The Schrieffer-Wolff transformation
 \begin{equation}
      \hat U =\exp\left(i \frac{k}{k_c}\frac{\sqrt{\omega}}{2\sqrt{\Omega}}(\hat a +\hat a^\dagger){\hat \sigma_y} \right), 
 \end{equation}
 is a simultaneous rotation of the spin by an angle $\phi = \frac{k}{k_c}\frac{\sqrt{\omega}}{\sqrt{\Omega}} \left(\hat a^\dagger + \hat a\right)$ and displacement of the harmonic oscillator by $\alpha = i\frac{k}{k_c}\frac{\sqrt{\omega}}{2\sqrt{\Omega}} \hat \sigma_y$. Applying this transformation to the Hamiltonian~\eqref{eq:qrform} yields
 \begin{align}
     \hat H =   \omega  \left(\hat a^\dagger-\alpha\right) \left(\hat a+\alpha\right) + \frac{\Omega}{2}  \left(\cos \phi \hat\sigma_z -\sin \phi \hat \sigma_x\right) + k\sqrt{\frac{ \omega}{2m}} \left(\hat a  + \hat a^\dagger\right) \left(\cos \phi \hat \sigma_x + \sin \phi \hat \sigma_z\right).
 \end{align}
 Assuming $\omega/\Omega \rightarrow 0$, we obtain Hamiltonian~\eqref{eq:effsocbec}
 \begin{align}
  \hat H \approx \omega \hat a^\dagger \hat a + \frac{\Omega}{2} \hat \sigma_z + k^2{\frac{ \omega}{2m \Omega}} (\hat a +  \hat a^\dagger)^2 \hat \sigma_z.
\end{align}
 \begin{figure}[htb!]
    \centering
    \includegraphics[width=\textwidth]{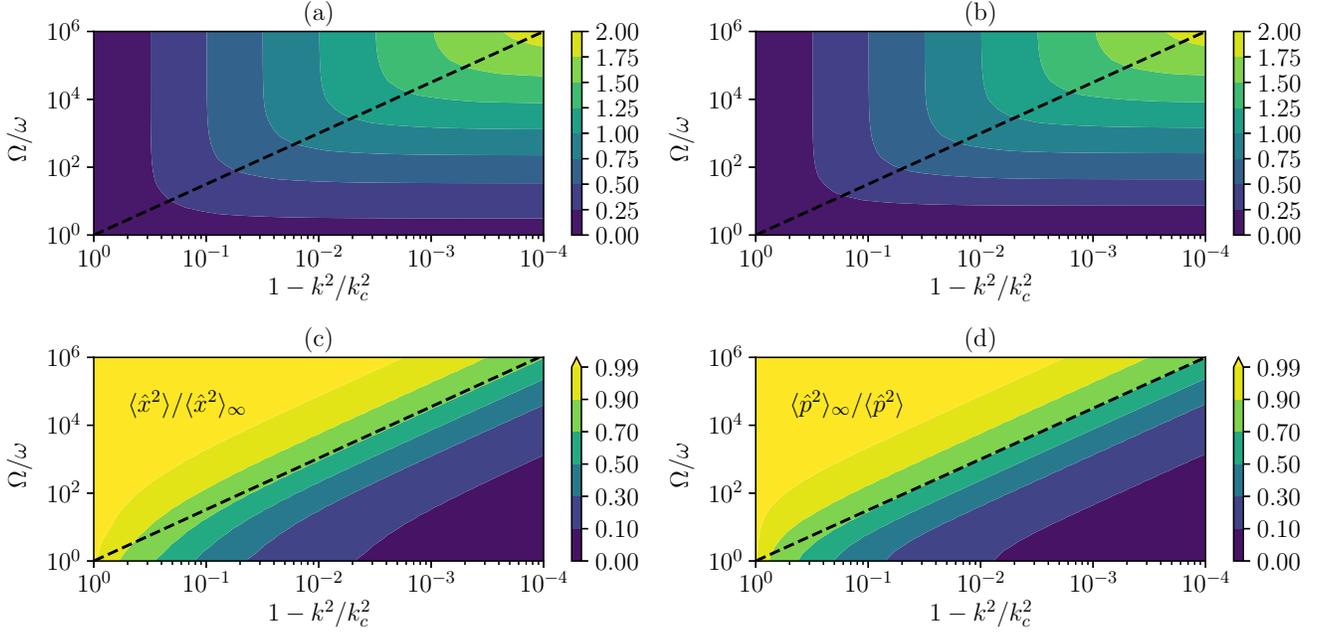}
    \caption{Comparison between the Hamiltonian~\eqref{eq:qrform} and~\eqref{eq:effsocbec}. For $(\omega/\Omega)^{2/3} \ll 1-k^2/k_c^2 $, the two Hamiltonians have the same ground state (the difference in negligible). Once this condition is not satisfied, the ground state of Hamiltonian~\eqref{eq:qrform} does not change anymore by further increasing $k$ towards $k_c$. (a) and (b) shows logarithm of $2 m \omega\langle \hat x^2 \rangle$ and $m \omega /2\langle \hat p^2 \rangle$. (c) and (d) shows second moments of $\hat x$ and $\hat p$ for Hamiltonian~\eqref{eq:qrform} normalized to second moments of $\hat x$ and $\hat p$ for Hamiltonian~\eqref{eq:effsocbec}.}
    \label{fig:fig3}
\end{figure}
By projecting the above Hamiltonian onto the spin down sector we finally get
 \begin{align}
  \hat H = \omega \hat a^\dagger \hat a - k^2{\frac{ \omega}{2m \Omega}} (\hat a +  \hat a^\dagger)^2,
 \end{align}
 which can be rewritten using the position and momentum operators in the form of a harmonic oscillator with a tunable frequency.
 \begin{equation}
       \hat H = \frac{p^2}{2m} + \frac{m \omega^2}{2}\left(1-\frac{k^2}{k^2_c}\right)\hat x^2.
 \end{equation}
 We now expect that  there exists a weaker condition than $\omega/\Omega \rightarrow 0$ that relates $\omega/\Omega$ and $k/k_c$ such that the effective description is still valid. In order to find this condition and check the limits of applicability of the effective Hamiltonian we now compute its ground state as a function of $1-k^2/k_c^2$ and $\Omega/\omega$ and compare with the ground state of the original Hamiltonian. We choose to compare the second moments of $\hat x$ and $\hat p$ as they characterize a Gaussian state centered at the origin of the phase space. The results of the simulations are presented in Fig.~\ref{fig:fig3} where we also indicate the $ 1-k^2/k_c^2 = ({\omega}/{\Omega})^{2/3}$ condition with a dashed-black line. Similar calculations can be performed for the Coloumb crystal.
 
  %%%%%%%%%%%%%%%%%%%%%%%%%%%%%%%%%%%%%%%%%%%%%%%%%%%%%%%%%%%%%%%%%%%%%%%%%%%%
%%%%%%%%%%%%%%%%%%%%%%%%%%%%%%%% APPENDIX B %%%%%%%%%%%%%%%%%%%%%%%%%%%%%%%%%
%%%%%%%%%%%%%%%%%%%%%%%%%%%%%%%%%%%%%%%%%%%%%%%%%%%%%%%%%%%%%%%%%%%%%%%%%%%%

\section{Effective Hamiltonian for the Coulomb crystal}\label{app:effcoucrys}
\begin{figure}[htb!]
    \centering
    \includegraphics[width=\textwidth]{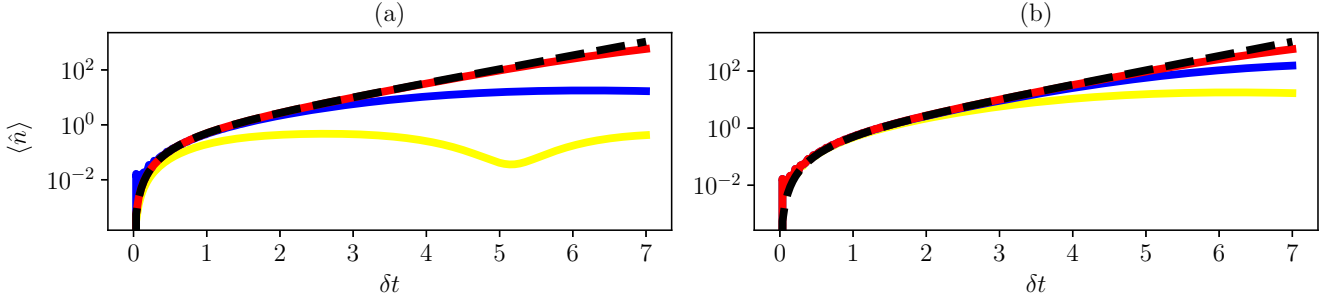}
    \caption{The dashed black line is the inverted harmonic oscillator dynamics. In (a) the number of ions is fixed to $N=1$; yellow, blue, and red line is a simulation for $N B/\delta = 0.1875$, $N B/\delta = 18.75$, and $N B/\delta = 1875$, respectively. In (b) the $B/\delta$ ratio is fixed to $B/\delta=18.75$; yellow, blue, and red line is a simulation for $N B/\delta = 18.75$ ($N=1$), $N B/\delta = 187.5$ ($N=10$), and $N B/\delta = 1875$ ($N=100$), respectively.}
    \label{fig:fig4}
\end{figure}

 Here we derive the effective Hamiltonian~\eqref{eq:cryseff}. Similarly as  in the case of a spin-orbit-coupled Bose-Einstein condensate, the Schrieffer-Wolff transformation
\begin{equation}
    \hat U =  \exp\left(i \frac{B_c}{B }\frac{\delta}{g\sqrt{N}}(\hat a +\hat a^\dagger){\hat S_y} \right)
\end{equation}
is a rotation of the collective spin around the $y$ axis by an angle $\varphi = \frac{B_c}{B }\frac{\delta}{g\sqrt{N}}(\hat a +\hat a^\dagger) $ and displacement of the harmonic oscillator by $\beta =  i \frac{B_c}{B }\frac{\delta}{g\sqrt{N}}{\hat S_y}$. Applying the transformation to Hamiltonian~\eqref{eq:culcrys} yields
\begin{equation}
    \hat H = - \delta \left(\hat a^\dagger-\beta\right)\left(\hat a +\beta\right)+  B\left(\cos \varphi \hat S_x + \sin \varphi \hat S_z\right) -  \frac{g}{\sqrt{N}}\left(\hat a + \hat a^\dagger\right)\left(\cos \varphi \hat S_z - \sin \varphi \hat S_x\right).
\end{equation}
Assuming $\delta/B \rightarrow 0$ we get
\begin{equation}
    \hat H  = - \delta \hat a^\dagger \hat a + B\hat S_x -  \frac{g^2}{2 N B}(\hat a + \hat a^\dagger)^2\hat S_x,
\end{equation}
which upon projecting on the low energy sector and using quadrature operators gives
\begin{equation}
    \hat H = \frac{\delta}{2}\hat P^2 + \frac{\delta}{2}\left(1-\frac{B_c}{B}\right)\hat X^2,
\end{equation}
and describes a harmonic oscillator with a tunable frequency which can become negative. By quenching from paramagnetic to ferromagnetic phase this Hamiltonian will give a proper description until the number of excitations approaches the number of excitations characterizing the ground state of the quenched Hamiltonian which can be calculated to be (assuming $B<B_c$)
\begin{equation}
    \langle \hat a^\dagger \hat a\rangle_{\mathrm{GS}} = \frac{N B}{4 \delta}\left(\frac{B_c}{B }-\frac{B}{B_c}\right),
\end{equation}
where the subscript GS indicates the ground state. Therefore, for a fixed $B$ the time for which the effective description is valid can be extended by increasing $\frac{N B}{\delta}$. This is illustrated in Fig.~\ref{fig:fig4} where we plot the number of excitations as a function of time for different parameters.

 %%%%%%%%%%%%%%%%%%%%%%%%%%%%%%%%%%%%%%%%%%%%%%%%%%%%%%%%%%%%%%%%%%%%%%%%%%%%
%%%%%%%%%%%%%%%%%%%%%%%%%%%%%%%% APPENDIX C %%%%%%%%%%%%%%%%%%%%%%%%%%%%%%%%%
%%%%%%%%%%%%%%%%%%%%%%%%%%%%%%%%%%%%%%%%%%%%%%%%%%%%%%%%%%%%%%%%%%%%%%%%%%%%

\section{Quantum Fisher information for a sudden quench in a Coulomb crystal}\label{app:qficrys}
Here we calculate the quantum Fisher information for a sudden quench from the paramagnetic phase to the ferromagnetic phase. The quantum Fisher information can be calculated as a variance of the local generator $\hat h_\lambda$, where $\lambda$ is the unknown parameter. The local generator $\hat h_\lambda$ for a Hamiltonian $\hat H = \hat H_0 +\lambda \hat H_\lambda$ can be expressed as
\begin{equation}
     \begin{split}
       \hat h_\lambda =  i e^{i \hat H t} \left(\partial_\lambda e^{-i \hat  H t}\right) = \int_0^t   e^{i \hat H s} \hat H_\lambda  e^{-i \hat  H s} \mathrm{d}s  =  \int_0^t \sum_{n=0}^\infty \frac{(is)^n}{n!}\left[\hat H,\hat H_\lambda \right]_n \mathrm{d}s
       =  - i \sum_{n=0}^\infty \frac{(it)^{n+1}}{(n+1)!}\left[\hat H,\hat H_\lambda \right]_n,
     \end{split}
 \end{equation}
 where $[\hat H,\hat H_\lambda ]_{n+1} =[\hat H, [\hat H,\hat H_\lambda ]]_{n}$ and $[\hat H,\hat H_\lambda]_{0} = \hat H_\lambda$. Now it can be shown that if $[\hat H,\hat H_\lambda] = i \hat C$, $[\hat H, i \hat C] = -\hat D$, and $[\hat H,- \hat D] = i \epsilon \hat C$, then
 \begin{equation}
         [\hat H,\hat H_\lambda ]_{2n+1} = i \epsilon^n \hat C \quad \mathrm{ and } \quad
         [\hat H,\hat H_\lambda ]_{2n+2} = - \epsilon^n \hat D. 
 \end{equation}
Therefore, one can express the local generator $\hat h_\lambda$ as
\begin{equation}
\hat h_\lambda = \hat H_\lambda t - i \sum_{n=0}^\infty \frac{(it)^{2n+2}}{(2n+2)!}i\epsilon^n \hat C + i \sum_{n=0}^\infty \frac{(it)^{2n+3}}{(2n+3)!}\epsilon^n \hat D =\hat H_\lambda t +\frac{\cos(\sqrt{\epsilon}t)-1}{\epsilon}\hat C -\frac{\sin(\sqrt{\epsilon}t)-\sqrt{\epsilon} t}{\sqrt{\epsilon}\epsilon}\hat D.
\end{equation}
Taking the effective Hamiltonian for the Coulomb crystal
\begin{equation}
        \hat H = \frac{\delta}{2}\hat P^2 + \frac{\delta}{2}\left(1-\frac{B_c}{B}\right)\hat X^2,
\end{equation}
assuming that $\delta$ is unknown, choosing $\hat H_\delta = \frac{1}{2}(\hat X^2+\hat P^2)$ and $\hat H_0 = -\frac{g^2}{2 B} \hat X^2$, one can show that
\begin{equation}
\begin{split}
\hat h_\delta  = &\frac{t}{2}\left(\hat X^2+\hat P^2\right) -\frac{g^2}{2B }\frac{\cos\left(\sqrt{\epsilon} t\right)-1}{ \epsilon}\left(\hat X \hat P + \hat P \hat X\right) +\frac{g^2 \delta}{B }\frac{\sin\left(\sqrt{\epsilon} t\right)-\sqrt{\epsilon} t}{\epsilon^{3/2}}\left(\hat X^2-\hat P^2\right)  - \frac{g^4}{B^2}\frac{\sin\left(\sqrt{\epsilon}  t\right)-\sqrt{\epsilon}  t}{\epsilon^{3/2}}\hat X^2,
\end{split}
\end{equation}
where $\epsilon = 4\delta^2({1-B_c/B})$. Once $\epsilon$ becomes imaginary ($B<B_c$), the sine and cosine functions become hyperbolic which reflects the exponential generation of excitations following a quench and leads to exponential dependence of the quantum Fisher information $\mathcal{I}_\delta = 4\Delta^2\hat h_\delta$ on time. A similar calculation can be performed treating other parameters as unknown as well as for the spin-orbit-coupled Bose-Einstein condensate.

 %%%%%%%%%%%%%%%%%%%%%%%%%%%%%%%%%%%%%%%%%%%%%%%%%%%%%%%%%%%%%%%%%%%%%%%%%%%%
%%%%%%%%%%%%%%%%%%%%%%%%%%%%%%%% APPENDIX D %%%%%%%%%%%%%%%%%%%%%%%%%%%%%%%%%
%%%%%%%%%%%%%%%%%%%%%%%%%%%%%%%%%%%%%%%%%%%%%%%%%%%%%%%%%%%%%%%%%%%%%%%%%%%%

\section{simulating homodyne detection}\label{app:homodyne}
Here we present simulations explaining how one can simulate homodyne detection of center-of-mass excitations. In order to simulate the homodyne detection, we decouple the center-of-mass mode from the spin degree of freedom after the final state is obtained. Waiting subsequently for a proper amount of time makes it possible to rotate the state in the phase space and align it with the real (position) or imaginary (momentum) axis. Therefore by measuring the distribution of position or momentum, one can measure an arbitrary quadrature. This mechanism is presented in Fig.~\ref{fig:fig3}, where we plot the Husimi representation $Q(\alpha) = \frac{1}{\pi}\langle \alpha|\hat \rho_\psi(t) |\alpha \rangle$ (where $|\alpha\rangle$ is a coherent state with amplitude $\alpha$ and $\hat \rho_\psi(t)\equiv |\psi(t)\rangle \langle \psi(t) |$ is the state of the system) for the (a) initial ($t=0$) and (b) final ($t=T$) state, as well as the rotated final state aligned with the (c) $\hat x$ and (d) $\hat p$ quadratures.

\begin{figure}[htb!]
    \centering
    \includegraphics[width=\textwidth]{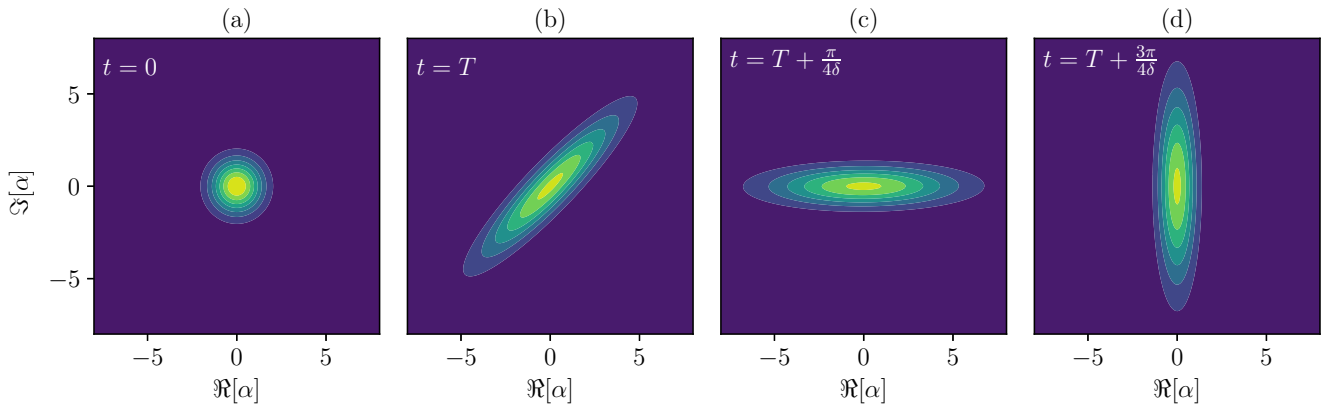}
    \caption{Simulating homodyne detection of center-of-mass excitations. (a) is the initial state (vacuum), (b) is the final state (squeezed state), (c) is the final state rotated to align with the $\hat x$ quadrature, (d) is the final state rotated to align with the $\hat y$ quadrature.}
    \label{fig:homodyne}
\end{figure}

 %%%%%%%%%%%%%%%%%%%%%%%%%%%%%%%%%%%%%%%%%%%%%%%%%%%%%%%%%%%%%%%%%%%%%%%%%%%%
%%%%%%%%%%%%%%%%%%%%%%%%%%%%%%%% APPENDIX E %%%%%%%%%%%%%%%%%%%%%%%%%%%%%%%%%
%%%%%%%%%%%%%%%%%%%%%%%%%%%%%%%%%%%%%%%%%%%%%%%%%%%%%%%%%%%%%%%%%%%%%%%%%%%%

\section{Finite temperature calculations}\label{app:finiteTBEC}
\begin{figure}[htb!]
    \centering
    \includegraphics[width=\textwidth]{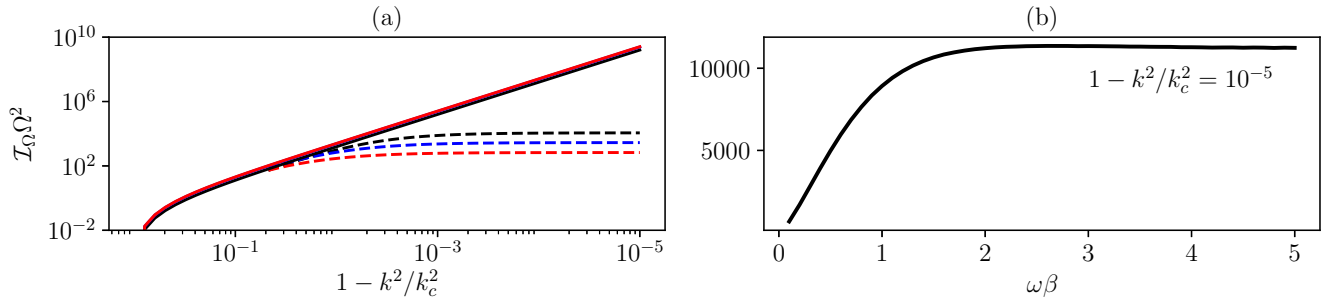}
    \caption{Quantum Fisher information for various temperatures of a spin-orbit coupled Bose-Einstein condensate. In panel (a) the solid lines represent $\Omega/\omega \rightarrow \infty$ case and the dashed lines represent $\Omega/\omega = 10^4$ case. Black, blue, and red colors correspond to $\omega\beta = 2,0.3,0.1$, respectively. In panel (b), we plot the final value of the quantum Fisher information as a function of $\omega \beta$. Once $\omega \lesssim 2$ the plateau value of the quantum Fisher information starts to decrease.}
    \label{fig:socemetT}
\end{figure}

In the main text, we have focused on the zero temperature case where the system resides in its ground state and is described by a pure state. From a practical point of view, however, it is relevant to consider a finite temperature case where the system occupies (incoherently) excited levels as well. In such a case, the state of the system can be described using a density matrix operator (statistical mixture)
\begin{equation}
    \hat \rho = \sum_{n=0}^{\infty} p_n |n\rangle \langle n|,
\end{equation}
where $n$ labels the excited states and $p_n$ is given by
\begin{equation}
    p_n = \frac{1}{Z}\exp(- \beta E_n),
\end{equation}
where $Z = \sum_n \exp(- \beta E_n)$ is the partition function (statistical sum) with $E_n$ being the energy of the $n$th excited state, and $\beta = 1/k_b T$ where $k_b$ is the Boltzmann constant and $T$ is the temperature. In order to calculate the quantum Fisher information for a statistical mixture, we have to use the following formula
\begin{equation}
    \mathcal{I_\lambda} = 2 \sum_{n,m}\frac{(p_n-p_m)^2}{p_n+p_m}|\langle n| \hat h_\lambda | m \rangle |^2.
\end{equation}
The results of numerical simulations for the spin-orbit coupled Bose-Einstein condensate are presented in Fig~\ref{fig:socemetT}. In panel (a), we plot the quantum Fisher information for the limiting case where $\omega/\Omega \rightarrow 0$ (solid lines) and for finite $\omega/\Omega$ (dashed lines). Interestingly, the quantum Fisher information increases with growing temperature until the maximum amount of squeezing is obtained. Once the plateau is reached, increasing the temperature decreases the quantum Fisher information. This is further illustrated in panel (b), where we plot the quantum Fisher information as a function of temperature for the case where the maximum amount of squeezing has been reached ($1-k^2/k_c^2 = 10^{-5}$). The (initial) growth of the quantum Fisher information is caused by the increased number of center-of-mass excitations for a thermal state.

\begin{figure}[htb!]
    \centering
    \includegraphics[width=\textwidth]{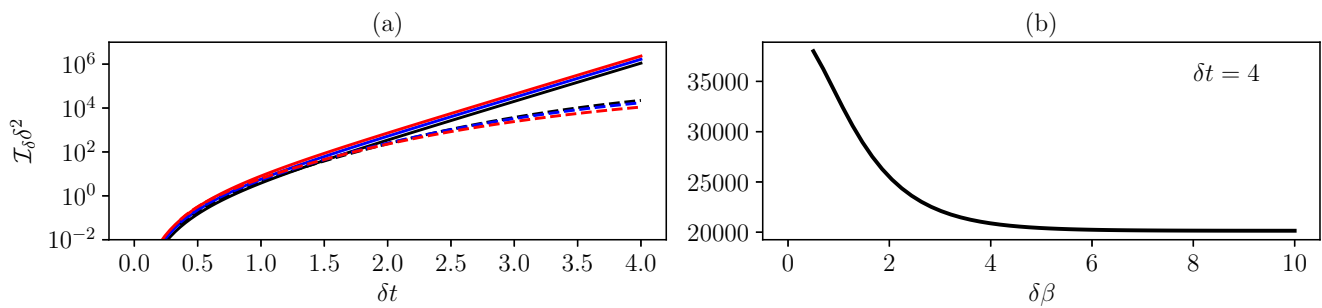}
    \caption{Quantum Fisher information for various temperatures of a Coulomb crystal. In panel (a) the solid lines represent $N B/\delta \rightarrow \infty$ case and the dashed lines represent $N B/\delta = 200$ case. Black, blue, and red colors correspond to $\delta\beta = 2,1,0.5$, respectively. In panel (b), we plot the value of the quantum Fisher information for the infinite case ($N B/\delta \rightarrow \infty$) as a function of $\delta\beta$ for $\delta t = 4$. Once $\delta \beta \lesssim 4$, the final value of the quantum Fisher information starts to increase.}
    \label{fig:crysT}
\end{figure}

The results of numerical simulations for the Coulomb crystal are presented in Fig~\ref{fig:crysT}. Similarly as in Fig.~\ref{fig:socemetT}, in panel (a) we plot the quantum Fisher information for the limiting case where $NB/\delta \rightarrow 0$ (solid lines) and for finite $N B/\delta$ (dashed lines). The quantum Fisher information increases (again) with growing temperature until the inverted harmonic oscillator picture is no longer applicable. Once quantum Fisher information does not grow exponentially anymore, the effect of increased temperature decreases the quantum Fisher information. The growth of the quantum Fisher information is further illustrated in panel (b), where we plot the quantum Fisher information as a function of temperature for the case when the inverted harmonic oscillator picture is still valid (without the loss of generality, we choose $\delta t = 4$ point).

% %%%%%%%%%%%%%%%%%%%%%%%%%%%%%%%%%%%%%%%%%%%%%%%%%%%%%%%%%%%%%%%%%%%%%%%%%%%%
%%%%%%%%%%%%%%%%%%%%%%%%%%%%%%%%%  BIB  %%%%%%%%%%%%%%%%%%%%%%%%%%%%%%%%%%%%
%%%%%%%%%%%%%%%%%%%%%%%%%%%%%%%%%%%%%%%%%%%%%%%%%%%%%%%%%%%%%%%%%%%%%%%%%%%%
\twocolumngrid

\end{document}